\documentclass[runningheads]{llncs}
\usepackage[T1]{fontenc}
\usepackage{wrapfig}
\usepackage{subcaption}
\usepackage{comment}
\usepackage{xcolor}
\usepackage{graphicx}

\usepackage{caption}

\begin{document}
\title{A Survey of AI Misconception and Adoption Among Indonesian K-12 Teachers}
\author{Alham Fikri Aji\inst{1,4}\orcidID{0000-0001-7665-0673} \and
Afifa Amriani\orcidID{0009-0009-9606-1206} \and
Rendi Chevi\inst{1}\orcidID{0000-0002-7917-6556}
\and
Ayu Purwarianti\inst{2}\orcidID{0000-0002-5016-3700}
\and
Derry Wijaya\inst{3,4}\orcidID{0000-0002-0848-4703}
}
\institute{Mohamed bin Zayed University of Artificial Intelligence, United Arab Emirates \\
\email{alham.fikri@mbzuai.ac.ae}
\and
Institut Teknologi Bandung, Indonesia 
\and
Boston University, USA
\and
Monash University, Indonesia\\
\email{derry.wijaya@monash.edu}
}

% Shortened title and authors
\titlerunning{AI Misconception and Adoption Among Indonesian K-12 Teachers}
\authorrunning{A.F. Aji et al.}

\maketitle              % typeset the header of the contribution
\begin{abstract}
Indonesia’s fragmented geography and uneven infrastructure pose persistent challenges for adopting new educational technologies. While national initiatives aim to mainstream AI in K–12 schools by 2025, policy readiness does not automatically mean teachers are ready for the classroom. This study surveys Indonesian K–12 teachers to examine: (1) their understanding of AI, (2) how they and their students use AI, and (3) their views on the acceptability of AI in learning activities. Through semi-automated thematic analysis, we identify common misconceptions and use cases. Our findings reveal a notable mismatch: while teachers find AI use in education acceptable and support it for the sake of efficiency, they remain fully aware of the risks to learning depth and academic honesty. We also find that this high level of acceptance is currently coupled with a poor understanding of how AI actually works.

\keywords{AI Misconception \and LLMs \and Pedagogy \and Indonesia.}
\end{abstract}
\section{Introduction}

Global educational landscape is undergoing a paradigm shift driven by the rapid adoption of AI technologies. International guidelines and policy recommendations increasingly stress that realizing AI’s educational benefits depends on system readiness, particularly, the teacher's capacity, equitable infrastructure, and governance~\cite{funa2025policy,pedro2019artificial}. However, in many middle-income and structurally diverse education systems, AI adoption is advancing on ambitious policy timelines while teacher preparation and other enabling conditions remain uneven across regions and school sectors~\cite{landa2025artificial}. This mismatch creates a key implementation risk: self-reported "readiness" may not correspond to classroom-relevant understanding and judgment, limiting the ability to target professional development effectively or anticipate downstream pedagogical harms. Moreover, without sustained support for teacher AI literacy, gaps in preparedness can widen existing digital divides rather than reduce them~\cite{daher2025integrating}. In this study, we refer to AI literacy as the teachers' demonstrated conceptual understanding of what AI systems, including Large Language Models (LLM)-based tools, can and cannot do, and their ability to make informed pedagogical decisions about how AI should be used in classroom practice.

Indonesia provides a salient case of these broader dynamics. The transition is being accelerated by aggressive government mandates, including the National Strategy for Artificial Intelligence 2020–2045~\cite{KORIKA2024} and the 2025 Academic Draft on Coding and AI Education~\cite{Kemendikdasmen2024}. These policy frameworks seek to integrate AI and coding into the national curriculum beginning in the 2025–2026 academic year. As this deadline approaches, a critical question remains: \textit{Are Indonesian teachers prepared for this pedagogical shift?}

Indonesia’s geographical  and socioeconomic diversity pose structural barriers to uniform technology adoption. Infrastructure gaps constrain national readiness: approximately 65\% of schools reportedly lack stable internet and 35\% operate without reliable electricity~\cite{badaruddin2025digital}. Systemic disparities further influence teacher capacity, as urban educators typically access better compensation and professional development than those in rural regions~\cite{Dini2024}. Consequently, opportunities to understand and ethically evaluate AI are likely uneven across regions.

As AI adoption intensifies, it remains unclear how teachers will navigate complex pedagogical decisions given uncertainties regarding their own AI conception and literacy. Research highlights risks of over-reliance, including reduced critical thinking~\cite{kosmyna2025brainchatgptaccumulationcognitive,mollick2023assigning,shen2026ai}. Because teachers’ conceptions determine how tools are operationalized, a lack of understanding regarding AI capabilities may reinforce passive learning rather than critical engagement. To address this, we survey Indonesian educators’ readiness for AI adoption. While utilizing Indonesian data, the measurement approach and readiness tensions identified remain relevant to other systems facing similar constraints. The preliminary study is guided by the following research questions:

\vspace{1mm}
\noindent\textbf{RQ.1}: What (mis)conceptions do K-12 Indonesian teachers have towards AI?

\vspace{1mm}
\noindent\textbf{RQ.2}: What are the practical use cases of AI among K-12 Indonesian teachers?

\vspace{1mm}
\noindent\textbf{RQ.3}: How do teacher perceive associated benefits and risks of AI in education?

\vspace{1mm}
\noindent\textbf{RQ.4}: What are the teachers' pedagogical stances regarding the acceptability of student AI adoption in learning activities?

% \begin{itemize}
% \item \textbf{RQ.1}: What are the (mis)conceptions the K-12 Indonesian teachers have towards AI?
% \item \textbf{RQ.2}: What are the practical use cases of AI among K-12 Indonesian teachers, and how do they perceive its associated benefits and risks?
% \item \textbf{RQ.3}: What are the teachers' pedagogical stances regarding the acceptability of student AI adoption in learning activities?
% \end{itemize}

\section{Background and Related Works}

Indonesia's diverse K-12 system is governed by a national framework shaping institutional provision and teacher support. The National Education System Law (Law No. 20/2003)~\cite{indonesia2003_sisdiknas} permits multiple provider types (public and private), creating variation in curricula and resourcing. Teacher roles also vary; Law No. 14/2005~\cite{indonesia2005_gurudosen} defines teachers as professional educators, yet employment ranges from state positions (ASN) to non-ASN hires, impacting incentives and professional development access~\cite{indonesia2023_asn}. Despite national funding like the School Operational Assistance grant (BOS), infrastructure differs by locality, affecting technological readiness. These structural differences create a measurement challenge: since opportunities to practice AI pedagogy are uneven, assessment must distinguish perceived readiness from classroom-relevant competence.

Against this backdrop, the government has begun scaling teacher capacity in AI. In 2025, the Ministry reported deploying 2,915 facilitators to train 62,898 teachers across 53,568 schools. While an important step, this initiative underscores the gap between policy ambition and scale: Indonesia has 3,857,718 K–12 teachers across 276,415 schools. Given this size, accurately measuring AI literacy is essential for targeting professional development and monitoring whether training translates into competence.

Most evidence on Indonesian teachers' AI literacy relies on self-report instruments, which capture perceptions but are limited proxies for competence. Self-reports are vulnerable to bias and inconsistent interpretations, particularly given uneven AI exposure. Recent studies exemplify this. Drajati et al. (2025)~\cite{DRAJATI25} surveyed 176 EFL teachers using translated Artificial Intelligence Literacy Scale (AILS) items~\cite{WANG23} and interviews; they found teachers least confident in using AI but confident in recognizing misuse, alongside resource constraints. Similarly, Batubara et al. (2025)~\cite{BATUBARA25} examined elementary teachers’ readiness using adapted self-efficacy scales~\cite{WANG23}~\cite{DOMINGUEZ25} and open-ended responses, reporting high perceptions but lower practical readiness, with concerns regarding privacy and bias.

Prior studies provide useful signals about teachers' perceived readiness for AI, but offer less insight into how teachers articulate their understanding of AI and its classroom use. Under pressure to adopt AI, fixed-response survey answers may reflect aspiration rather than how teachers actually describe AI, reported uses, and acceptable classroom boundaries. Our study addresses this gap by using open-ended items and task-specific pedagogical judgments to elicit how teachers understand AI concepts and justify classroom uses. This approach allows us to examine the substance of teachers' expressed understanding, not only their confidence ratings. These findings provide a foundation for designing differentiated professional development, evaluating training efficacy, and helping policymakers target support.

\section{Methods}

\subsection{Participants Recruitment}

% Our participants comprised 132 K-12 Indonesian teachers. To ensure reasonable representation across different school levels and regions, we employed quota sampling during recruitment of our participants. 

% Quota sampling strategy
% Distribution

We recruited participants through our professional word-of-mouth networks, reaching out directly to K–12 schools (elementary, junior high, and high school) across Indonesia to build an initial pool of potential teacher participants. At each school, a contact person circulated a study advertisement and an online registration via Google Form through which teachers could submit their interest in the study. The form briefly describe that we are recruiting teachers to study how AI being adopted within their teaching-learning activities and they are being asked to fill in several demographic informations: sex, age, subject they taught, education level, employment status, school’s sector, and region.

To ensure more balanced representation across different socioeconomic backgrounds of teachers, we employed quota sampling when screening participants from the pool. Specifically, we set quotas for sex, school level, school region, and employment status. However, because the pool of registrants from outside West Java was limited, we did not apply quota constraints for that subgroup, instead, we invited participants from outside West Java via random sampling from the available pool. Screened participants were then invited to the study and received IDR 50,000 compensation for an approximately one-hour session, which is twice the average hourly rate in Indonesia.

% The participants of this study comprised 132 teachers recruited from various regions in Indonesia. The sample was geographically distributed, with 61.4\% (n=81) of the participants in West Java and 38.6\% (n=51) in regions outside of Java (Outer Java).

% The study information was shared directly with schools and through professional word-of-mouth networks. Interested participants registered through an online platform. To ensure a balanced representation of the teaching population, a quota sampling technique was applied based on three primary criteria: education level taught (Primary, Middle and Secondary), geographic location and employment status (Permanent or Contract). The eligible participants were then contacted to complete the consent form and the online questionnaire. Approximately 73\% of the respondents taught in public schools, while 27\% taught in private schools. Among all respondents, 77 (58.8\%) were female and 54 (41.3\%) were male, with an age range of 19-56.

% The participants represented a diverse range of educational backgrounds. The majority of participants had a bachelor’s degree (64\%), while a smaller number had postgraduate qualifications (26\%), a limited number had secondary education (10\%) and one person had a Diploma. The majority of the teachers taught Science, Technology and Math subjects.

\subsection{Survey Design}

We designed a mixed-format survey of 13 items spanning open-ended responses, multiple-response selections, and Likert-style ratings, divided into 3 sections. These sections aimed to probe teachers' AI (mis)conceptions and literacy (Q1--Q4), reported AI use in teaching-learning activities (Q5--Q12), and pedagogical judgments about the acceptability of students' AI usage in learning activities (Q13). Our survey items are in\footnote{\label{fn:artefact}We attached our survey items, prompts, rubrics, and other artefacts here: \url{https://github.com/afaji/AIED2026-Indonesian-AI-misconception}}.

\vspace{2mm}

\noindent\textbf{Survey Items Rationale}\quad For Q1-Q4, we adapted and localized 3 open-ended questions from~\cite{MERTALA2024100630} that aimed to elicit the teacher’s initial conceptions of AI in a minimally leading way. We further add a new question (Q2) to explicitly probe the teacher's mental models how popular LLM works. In Q5–Q6, we elicited concrete, self-reported examples of how teachers and students use AI in everyday learning contexts. Q7–Q8 then captured teachers’ stance toward AI use by teachers versus students, along with the reasons underlying each stance as well as the conditions under which teachers would change their position. In Q9–Q12, we asked teachers to articulate perceived positive and negative impacts of AI use for teachers and students. Finally, we probed teachers’ pedagogical decision about when student AI use is acceptable by eliciting task and stage-specific judgments rather than general attitudes. In Q13, teachers indicated which learning tasks they believe students use AI for. For each selected task, we asked teachers to rate the appropriateness of student AI use on a structured scale and to further evaluate AI use across fine-grained stages within that task. This branching, stage-level design was intended to probe teachers’ boundaries of AI usage in a format that supports comparison across various learning activities.

\vspace{2mm}

\noindent\textbf{Refining Survey Items in Pilot Study}\quad We conducted a pilot study with six teachers recruited through our professional networks to validate and refine our initial questionnaire items, particularly those newly developed (Q5–Q13). After the teachers completed the survey, we invited them for an unstructured interview, providing feedback that informed several revisions. Notably, Q3 was rephrased after teachers expressed confusion and felt it was assessing their technical knowledge of AI, it was rephrased to focus on their own understanding of what happened after one gave inputs to the AI. Discussions with teachers on their AI use in teaching and learning also strongly resonated with Q5–Q12, sparking engaging discussion between teachers and researchers, and confirming the items relevance and ability to capture diverse perspectives on teacher’s AI use. Finally, Q13’s nested, mixed-format design was refined based on insights that teachers’ attitudes toward AI wildly vary across different learning activities, leading us to specify activity types more explicitly to better capture their stances.

% The four task types (short-answer tasks, literacy/numeracy assignments, essays, and projects) were chosen as illustrative classroom categories that span shorter and more extended student work products commonly recognized across K--12 settings. These categories were refined through teacher consultation following our initial pilot study and were not intended as an exhaustive taxonomy. \textcolor{red}{TODO Pilot}

%Data collection
%From August to September 2025, we performed data collection. We conduct online. First is Java Next is Non Java

%- attention check
%- literacy score
%- pedagogy agreement score

\subsection{Data Collection and Analysis}

%From August to September 2025, we performed data collection. We conduct online. First is Java Next is Non Java

We conducted the data collection from September to November 2025. Each screened participant received a study invitation link, directing them to our online survey. Survey responses were stored in a tabular format, where each row corresponds to a participant and columns include: (1) item-level responses (open-ended text responses and structured selections and Likert ratings) and (2) demographic information, yielding a mixed-type tabular data.

% \paragraph{Analysis Overview}

\vspace{2mm}

\noindent \textbf{Analysis Overview}\quad We analyzed the dataset using a mixed-methods workflow. For structured items, we summarized the responses descriptively (e.g., frequencies and distributions) and used these variables to contextualize patterns in the qualitative findings. For open-ended items, we applied a human-in-the-loop NLP workflow to identify and label the major themes (topics) participants expressed, combining topic modeling with researcher review and LLM-assisted rubric development. 

% \paragraph{Human-in-the-Loop NLP Pipeline}
\vspace{2mm}

\noindent\textbf{Human-in-the-Loop NLP Workflow}\quad Following prior work on LLM-assisted topic modeling~\cite{wang2023prompting}, we analyzed each open-ended question using a three-step workflow: (1) \textit{topic generation} with Latent Dirichlet Allocation (LDA)~\cite{blei2003latent}, (2) \textit{topic refinement} (collapse/split) with LLM support and human review~\cite{wang2023prompting}, and (3) \textit{topic representation} via an LLM-generated coding rubric~\cite{hashemi-etal-2024-llm}. We first ran LDA with 2--10 topics per question and selected the topic count using the $C_v$ coherence score~\cite{10.1145/2684822.2685324}. We then reviewed the 20 most representative responses per topic to draft provisional labels. Next, ChatGPT (GPT-5.2) was used in a human-in-the-loop process to improve interpretability by proposing and validating merges/splits. After finalizing the topic set, ChatGPT generated operational definitions and a coding rubric for each label. Finally, GPT-5 mini applied the rubric at scale to assign a label to each response; researchers iteratively audited representative outputs and refined rubrics as needed for alignment. The average agreement rate between the LLM-assigned codes and human judgment, computed on 10\% sample of the data across questions, is 95.6\%; showing strong alignment between LLM and human judgments based on the rubric\footref{fn:artefact}.

\vspace{2mm}

\noindent \textbf{Filtering AI-generated Responses}\quad We hypothesize that AI-generated responses are the main concern threatening our data quality. Since automatic detectors are unreliable for short, non-English text~\cite{wang-etal-2024-m4}, we used a two-stage human-in-the-loop filter informed by evidence that experienced LLM users better recognize AI-written text~\cite{russell2025people}. We first used a RegEx-based exclusion-first heuristic to surface candidate cases. The script returned no alert when a response showed cues that we treated as more characteristic of human-written Indonesian survey text, including reduplication such as ``orang2'', abbreviations such as ``dll'', ``dst'', or ``dkk'', non-standard spacing around punctuation, lowercase list items, lowercase or variably capitalized forms of ``ai''/``ChatGPT'', lowercase sentence starts, or missing sentence-final punctuation. Only responses without those cues were then checked for red flags, namely length greater than 400 characters, double newlines, numbered or bulleted list structure, non-ASCII characters or emojis, the formal pronoun ``anda'', and parenthetical explanations such as ``Artificial Intelligence (AI)''. Two authors then independently reviewed flagged cases, and we excluded a response only when both judged it to be AI-generated.

\section{Result}

% In this section, we present our findings for each research question based on topic modeling analysis. We report the prevalence of each topic as a percentage rounded to one decimal place. The full anonymized survey dataset is available upon request.

\subsection{Distribution of Teacher Participants}

Total number of teachers we managed to reach out and completed the survey are 132, with 61.4\% (n=81) from West Java and 38.6\% (n=51) from outside of West Java. Due to quota sampling on West Java participants, we managed to gather decently balanced representation on sex (Male: 42.0\%, Female: 58.0\%), school level (Elementary: 30.8\%, Junior High: 30.8\%, High School: 38.0\%, employment status (Full-time: 56.7\%, Part-time: 43.2\%), but not on school sector (Public: 65.4\%, Private: 34.5\%). For outside West Java the distribution is as follows, sex (Male: 39.2\%, Female: 60.7\%), school level (Elementary: 43.1\%, Junior High: 11.7\%, High School: 45.0\%), employment status (Full-time: 88.2\%, Part-time: 11.7\%), school sector (Public: 90.1\%, Private: 9.8\%). We attach the complete demographic distribution in\footref{fn:artefact}. Among our 132 participants, 2 of them have been flagged as AI-generated and omitted from the final analysis.

% The participants of this study comprised 132 teachers recruited from various regions in Indonesia. The sample was geographically distributed, with 61.4\% (n=81) of the participants in West Java and 38.6\% (n=51) in regions outside of Java (Outer Java).

% The study information was shared directly with schools and through professional word-of-mouth networks. Interested participants registered through an online platform. To ensure a balanced representation of the teaching population, a quota sampling technique was applied based on three primary criteria: education level taught (Primary, Middle and Secondary), geographic location and employment status (Permanent or Contract). The eligible participants were then contacted to complete the consent form and the online questionnaire. Approximately 73\% of the respondents taught in public schools, while 27\% taught in private schools. Among all respondents, 77 (58.8\%) were female and 54 (41.3\%) were male, with an age range of 19-56.

% The participants represented a diverse range of educational backgrounds. The majority of participants had a bachelor’s degree (64\%), while a smaller number had postgraduate qualifications (26\%), a limited number had secondary education (10\%) and one person had a Diploma. The majority of the teachers taught Science, Technology and Math subjects.

\subsection{RQ.1: Conceptions of AI Among Indonesian Teachers}

%\begin{figure}[ht!]
%     \centering
%     \begin{subfigure}[b]{0.48\textwidth}
%         \centering
%         \includegraphics[width=\linewidth]{image/Q1.png}
%     \end{subfigure}
%     \hfill
%     \begin{subfigure}[b]{0.48\textwidth}
%         \centering
%         \includegraphics[width=\linewidth]{image/Q2.png}
%     \end{subfigure}
%     \caption{Distribution of topics on Q1 (left) and Q2 (right).}
%     \label{fig:q1-q2}    
%\end{figure}
To assess Indonesian teachers' conceptual understanding of AI, the survey began with two  items on general AI and LLMs. Responses to \textbf{Q1: ``Based on your current understanding, explain what is AI?''} clustered into three themes. \textbf{(T1) AI as a technology that mimics human} (55.4\%): many responses described AI as "artificial intelligence that mimics human intellectual abilities." [r116]; \textbf{(T2) AI defined through practical uses} (40\%): responses that emphasized utility, such as "applications that can make work easier" [r8]; and \textbf{(T3) AI framed primarily through risks, limitations, or governance as a key point} (4.6\%): a small minority focusing on harms or misuse, for example "AI that leads to criminal acts" [r45]. 

\begin{comment}
    \item \textbf{T1. A technology that mimics human (55.4\%).} Most answer follows similar idea such as "AI is artificial intelligence that mimics human intellectual abilities." (r116).
    \item \textbf{T2. Answer that focuses on the practical use (40\%)}, such as "applications that can make work easier" (r8).
    \item \textbf{T3. Answers focus on risks/limitations or governance as a key point (4.6\%).} For example "AI that leads to criminal acts" (r45).
\end{comment}

Most responses were high-level and lack depth, suggesting limited understanding of how AI works. To probe teachers' literacy about AI mechanisms more directly, we asked %We further explore teacher's literacy on how AI works by asking 
\textbf{Q2: ``Explain what ChatGPT does after receiving the instructions you type until it gives you an answer''}. Responses clustered into four themes: \textbf{(T1) High-level input-output interaction} (69.2\%), the majority of respondents describe ChatGPT as simply processing instructions and returning a response; their answers were often vague, demonstrated a lack of understanding regarding how AI functions beyond the immediate user interaction, and largely restated the question (e.g., "process and produce text according to the instructions given" [r13] or "send the answer back to the user via a conversational interface" [r21]. A small proportion used \textbf{(T2) Technical jargons} (4.6\%), mentioning concepts such as tokenization, generative AI, or decoding. Another substantial group reflected \textbf{(T3) Knowledge retrieval misconceptions} (21.5\%), suggesting that ChatGPT primarily searches the internet or a database for information (e.g., "looking for information from the data center he has" [r115]). Lastly, a minority emphasized \textbf{(T4) Personalization and adaptation} (4.6\%), focusing on the system's ability to learn user habits over time (e.g., "AI learns user habits" [r67]).

\begin{comment}
    \begin{itemize}
    \item \textbf{T1. High-level input-output interaction (69.2\%).} The majority of respondents describe the process simply as taking instructions and responding accordingly. These answers are often vague; for example, "process and produce text according to the instructions given" (r13) or "Send the answer back to the user via a conversational interface" (r21). Such responses demonstrate a lack of understanding regarding how AI functions beyond the immediate user interaction. Moreover, this answer rarely adds anything; as this is essentially rephrasing the questions.
    \item \textbf{T2. Technical jargons (4.6\%).} Some respondents provide technical details on LLM operations, mentioning concepts such as tokenization, generative AI, or the decoding process.
    \item \textbf{T3. Knowledge retrieval misconceptions (21.5\%).} Some respondents incorrectly suggest that LLMs like ChatGPT explore the internet or a database as a primary knowledge source, such as "Looking for information from the data center he has" (r115).
    \item \textbf{T4. Personalization and adaptation (4.6\%).} Finally, some focus on the system's ability to learn from the user, such as "AI learns user habits" (r67).
\end{itemize}
\end{comment}

Responses in topics \textbf{T1} and \textbf{T4} are typically superficial, failing to explore the actual mechanics of LLMs and instead focusing on the interaction itself. Those in \textbf{T3} are misinformed, viewing ChatGPT as a search engine or database retrieval system. Although the small number of respondents in \textbf{T2} utilize technical keywords, their explanations often remain high-level, such as simply stating that the system "uses NLP techniques" [r8]. Our findings suggest uneven and often limited AI literacy within this survey population.

To understand teachers' AI conception in terms of its perceived impacts and everyday uses, we asked \textbf{Q3: ``What are the reasons why people use AI in their personal and professional daily lives?''} Responses clustered into four themes. \textbf{(T1) Efficiency and productivity} (55.4\%) was the most prevalent, focusing on AI's ability to ``simplify and speed up work'' [r1] and make tasks ``practical, effective, and efficient'' [r6]. Some respondents noted that AI helps ``reduce the burden'' of work [r53], while a small number framed its use in terms of avoiding effort or seeking fast responses (e.g., being ``lazy to think'' [r5] or wanting ``quick answers'' [r82]). \textbf{(T2) Information retrieval} (25.4\%) framed AI as a sophisticated search tool for ``finding information and data according to needs'' [r3], identifying ``solutions to problems'' [r44], or obtaining quick references. \textbf{(T3) Content creation and ideation} (16.1\%) highlighted creative and teaching-related uses, such as ``making lesson plans, creating questions'' [r4], ``generating ideas or inspiration'' [r42], and helping``arrange words'' [r13] when respondents feel stuck. Finally, \textbf{(T4) Personal assistance and everyday needs} (3.1\%), a smaller group emphasized convenience-oriented functions of AI, including managing schedules as a ``virtual assistant'' [r48], providing ``entertainment and music recommendations'' [r58], or serving as a ``companion to share stories and seek advice'' [r61].
\begin{comment}
    \begin{itemize}
    \item \textbf{T1. Efficiency and productivity (55.4\%).} This is the most prevalent motivation, focusing on AI's ability to ``simplify and speed up work'' (r1) and make tasks ``practical, effective, and efficient'' (r6). Some respondents noted that it helps in ``reducing the burden'' of work (r53), while a few mentioned using it because they are ``lazy to think'' or want ``quick answers'' (r5, r82).
    \item \textbf{T2. Information retrieval (25.4\%).} Respondents in this category view AI as a sophisticated search tool. This includes ``finding information and data according to needs'' (r3), ``solutions to problems'' (r44), or quick references.
    \item \textbf{T3. Content creation and ideation (16.1\%).} Many respondents utilize AI for creative tasks, especially for teaching. This includes ``making lesson plans, creating questions'' (r4), ``generating ideas or inspiration'' (r42), and acting as a bridge to ``arrange words'' (r13) when they feel stuck.
    \item \textbf{T4. Personal assistant and daily life needs (3.1\%).} A smaller group highlights AI's role in personal convenience. This includes acting as a ``virtual assistant to manage schedules'' (r48), providing ``entertainment and music recommendations'' (r58), or serving as a ``companion to share stories and seek advice'' (r61).
\end{itemize}
\end{comment}

To gauge respondents' awareness of AI's broader societal uses, we asked \textbf{Q4: ``What do you think AI technology has been used for?''} Responses clustered into four themes. 
\textbf{(T1) Education \& Academic} (45.4\%) was most common, with respondents primarily identifying schools and office applications, such as producing ``teaching materials, quiz creation, and administrative reports'' [r48, r53], as well as supporting students with ``college assignments'' and summarizing study material [r108, r49]. \textbf{(T2) Creative \& Digital Content Production} (16.2\%) recognizing uses such as ``editing photos and videos'' [r29, r35], ``graphic design'' [r22], and generating ``social media content, slogans, or animations'' [r53, r101]. \textbf{(T3) Business \& Services} (10.8\%) captured consumer-facing and operational applications of AI, including ``online shopping and marketplaces'' [r1, r45], ``chatbot customer services'' [r101], ``mobile banking'' [r45], and ``marketing and business analytics'' [r49, r70]. Finally, in \textbf{(T4) Critical Sectors \& Governance} (27.7\%), a significant portion of respondents mentions AI use in high-stakes domains, including ``medical diagnosis and remote surgery'' [r49, r100, r104], ``defense-related systems like radar and drones'' [r7], and ``security applications such as facial recognition'' [r61, r104].

\begin{comment}
    \begin{itemize}
    \item \textbf{T1. Education \& Academic (45.4\%).} Respondents identify AI primarily for school and office functions, such as the creation of ``teaching materials, quiz creation, and administrative reports'' (r48, r53), as well as helping students with ``college assignments and summarizing study material'' (r108, r49).
    \item \textbf{T2. Creative \& Digital Content Production (16.2\%).} Many recognize AI's role in the creative industries, including ``editing photos and videos'' (r29, r35), ``graphic design'' (r22), and generating ``social media content, slogans, or animations'' (r53, r101).
    \item \textbf{T3. Business \& Services (10.8\%).} Respondents identify AI in consumer-facing and operational sectors, noting its use in ``online shopping and marketplaces'' (r1, r45), ``chatbot customer services'' (r101), ``mobile banking'' (r45), and ``marketing and business analytics'' (r49, r70).
    \item \textbf{T4. Critical Sectors \& Governance (27.7\%).} A significant portion of respondents mentions AI use in high-stakes domains. This includes applications in ``medical diagnosis and remote surgery'' (r49, r100, r104), ``defense systems like radar and drones'' (r7), and ``security features such as face recognition'' (r61, r104).
\end{itemize}
\end{comment}

\subsection{RQ.2: Practical Utility of AI in Education}

Overall, these responses suggest that teachers' familiarity with AI is strongly anchored in their professional domain, particularly within academic administration and pedagogy, even when asked about general use cases. We next examine how AI is integrated into teachers' day-to-day professional teaching activities. In response to \textbf{Q5: ``In your activities as a teacher, in what ways do you utilize AI?''}, the reported use cases clustered into four themes. \textbf{(T1) Teaching Planning and Design} (19.2\%) captured use cases for AI, particularly LLMs, for drafting or designing teaching plans, including ``teaching modules'' [r18], ``learning steps'' [r37], worksheets, and lesson personalization. \textbf{(T2) Generating Teaching Materials} (32.3\%), the most common response included the use of AI for producing teaching materials, such as slides, infographics, as well as ``editing image or videos'' [r15] and ``creating poster'' [r29]. \textbf{(T3) Grading and Assessment} (26.9\%) the second most common response captured the use of AI to aid in grading, whether by directly using AI to grade or by using AI to generate rubrics or answer sheets (e.g., ``automated grading'' [r131]). Finally, \textbf{(T4) Information Curation and Administrative Tasks (21.5\%).} described the use of AI as ``teaching tools'' [r76] to search for, summarize, translate, validate, simplify, or otherwise curate information to support teaching and related administrative work. 

\begin{comment}
    \begin{itemize}
    \item \textbf{T1. Teaching Planning and Design (19.2\%).} Some respondents utilize AI, and especially LLMs, for drafting or designing teaching plans, such as ``teaching modules'' (r18), ``learning steps'' (r37), worksheets, or lesson personalization.
    \item \textbf{T2. Generating Teaching Materials (32.3\%).} Some use AI to generate teaching materials, including slides, infographics, such as ``editing image or videos'' (r15) and ``create poster'' (r29)..
    \item \textbf{T3. Grading and Assessment (26.9\%).} Some respondents utilize AI to aid in grading, whether by directly using AI to grade or by using AI to generate rubrics or answer sheets, for instance ``automated grading'' (r131).
    \item \textbf{T4. Information Curation and Administrative Tasks (21.5\%).} Others use AI to aid in and curate information, such as for searching, summarizing, translating, validating, or simplifying information, or for ``teaching tools'' (r76).
\end{itemize}
\end{comment}

Finally, to examine teachers' awareness of students' AI use, we asked \textbf{Q6: ``In what ways do you think students utilize AI?''} Teachers' perceptions clustered into five themes. \textbf{(T1) No Use or No Knowledge} (9.2\%) captured, perhaps surprisingly, a small fraction of respondents who were unsure whether students used AI or believed they did not (e.g., ``unaware as it is not implemented yet'') [r15]. The most common theme was \textbf{(T2) Direct Answering}  (44.6\%), where teachers perceived the use of AI by students to obtain immediate answers for homework or assignments (e.g., ``looking for answer when given assignments'' [r117]). \textbf{(T3) Tutor-like support} (10.8\%), a minority of respondents described AI as a learning assistant used to ask questions, obtain explanations, and understand concepts (e.g., ``for learning reference sources, asking complex questions'' [r4]). \textbf{(T4) Information Retrieval} (19.2\%) reflected the use of AI to find or collect learning materials, information, or references (e.g., ``learning material quizzes'' [r54]). Finally, \textbf{(T5) Content Generation} (16.2\%) included the use of AI for producing slides, reports, or images (e.g., ``create poster'' [r29]).
\begin{comment}
    \begin{itemize}
    \item \textbf{T1. No Use or No Knowledge (9.2\%).} Perhaps surprisingly, a small fraction of respondents noted that they are unsure about the use of AI by students, or claimed that students do not use AI. For example, ``unaware as it is not implemented yet'' (r15).
    \item \textbf{T2. Direct Answering (44.6\%).} Many respondents noted the use of AI for directly answering questions in homework or assignments, for example ``looking for answer when given assignments'' (r117).
    \item \textbf{T3. Tutor (10.8\%).} Respondents noted AI usage by students as a learning assistant or tutor to help understand concepts, ``for learning reference sources, asking complex questions'' (r4), get explanations, practice, or clarify ideas.
    \item \textbf{T4. Information Retrieval (19.2\%).} AI is used mainly to find or collect learning materials, information, or references, for instance it used for ``learning material quizzes'' (r54).
    \item \textbf{T5. Content Generation (16.2\%).} Some respondents noted AI use for generating slides, reports, images, or other materials, for example ``create poster'' (r29).
\end{itemize}
\end{comment}

These findings indicate that AI is already embedded across multiple stages of the teaching pipeline, from lesson design to assessment. As a result, teacher support mechanisms need to evolve to reflect this reality. With the exception of a small minority, educators also recognize that students commonly use AI to obtain direct answers. Although many view this answer-seeking behavior unfavorably (as discussed in later sections), it does not appear to reduce their broader support for students’ use of AI.

\subsection{RQ.3: Percieved Benefit and Risk of AI Use in Education}

\subsubsection{Support for AI Use in Education} Next, we asked respondents' views on AI use in education. Specifically, we asked \textbf{Q7: ``Do you support the use of AI by teacher in learning activities?''} and \textbf{Q8: ``Do you support the use of AI by students in learning activities?''}. Majority of respondents expressed support, with 96.2\% supporting teachers' use of AI, and 84.6\% supporting students' use. To understand the basis for teachers' AI use, respondents who supported it were then asked \textbf{Q9a: ``I support the use of AI by teachers because..."}. Their explanations clustered into three themes. \textbf{(T1) Teacher Work Efficiency} (50.9\%), majority of respondents emphasized reduced workload and time savings, particularly for administrative tasks and the preparation of teaching materials (e.g., ``effectiveness; efficiency'' [r7]). Another common response was \textbf{(T2) Improvement of Student Learning} (43\%) that framed AI as enhancing the learning experience through innovation,  personalization, and more ``engaging learning experience'' [r131]. \textbf{(T3) Conditional Support} (6.1\%) expressed approval conditional upon specific safeguards, emphasizing that AI outputs must be verified and adapted to the pedagogical context (e.g., ``they must be filtered to suit the conditions of the students'' [r38]) and the need to avoid over-reliance. Overall, support for teachers' AI use centered on efficiency gains and personalization for improved learning experiences. 

\begin{comment}
    \begin{itemize}
    \item \textbf{T1. Teacher Work Efficiency (50.9\%).} Many respondents support AI because it reduces workloads by speeding up administrative tasks and the preparation of teaching materials, such as ``effectiveness; efficiency'' (r7) or other time-saving measures.
    \item \textbf{T2. Improvement of Student Learning (43\%).} Support in this category focuses on enhancing the learning experience through innovation and personalization, for instance, ``engaging learning experience'' (r131), where AI helps make lessons more adaptive to student needs.
    \item \textbf{T3. Conditional Support (6.1\%).} Some respondents provide support conditional upon certain guardrails, emphasizing that AI outputs must be verified and adapted to the pedagogical context, such as ``they must be filtered to suit the conditions of the students'' (r38) or the need to avoid over-reliance.
\end{itemize}

The overarching support for AI use by teachers centers on work efficiency, whether through the direct mention of speed or by assisting teachers with content production and reference searching. Personalization is also a key theme, as respondents believe it leads to improved learning experiences.
\end{comment}

Likewise, teachers who supported students' use of AI were asked \textbf{Q10a: ``I support the use of AI by students because..."}. Their responses clustered into six themes. \textbf{(T1) Information Access and Learning Resources} (27.3\%) emphasized AI's ability to provide fast access to references and broader knowledge (e.g., ``learning new things or content that is minimally covered at school.'' [r7]). \textbf{(T2) Understanding and Personalized Learning}  (30\%) favored the use of AI for explaining complex concepts and providing step-by-step guidance (e.g.,  ``understading task given'' [r31]). \textbf{(T3) Creativity and Idea Generation} (13.6\%) expressed support as AI can help students brainstorm and explore new perspectives (e.g., ``searching for creative for ideas'' [r117]). \textbf{(T4) Engagement and Interactive Experience} (9.1\%) described AI as making learning more interesting and interactive (e.g., ``learning becomes more fun, enjoyable, and smart.'' [r94]), boosting student motivation. \textbf{(T5) Efficiency and Productivity} (10.9\%) focused on completing tasks more quickly (e.g., ``speed up the completion of tasks'' [r8]). Finally, \textbf{(T6) Risks and Governance (9.1\%).} expressed support conditional on supervision and ethical safeguards to prevent misuse, including plagiarism or overdependence (e.g., ``because they generally or mostly use it to cheat'' [r2]). 
\begin{comment}
    \begin{itemize}
    \item \textbf{T1. Information Access and Learning Resources (27.3\%).} AI is supported for its ability to provide fast access to references and broaden students' knowledge, such as ``learning new things or content that is minimally covered at school.'' (r7).
    \item \textbf{T2. Understanding and Personalized Learning (30\%).} Some favor student use of AI because it helps explain complex concepts or provides step-by-step guidance, for instance ``understading task given'' (r31).
    \item \textbf{T3. Creativity and Idea Generation (13.6\%).} Some respondents express support because AI can help students in brainstorming and exploring new perspectives, such as ``searching for creative for ideas'' (r117).
    \item \textbf{T4. Engagement and Interactive Experience (9.1\%).} AI is seen by respondents as a way to make learning more interesting and interactive, with examples like ``learning becomes more fun, enjoyable, and smart.'' (r94), boosting student motivation.
    \item \textbf{T5. Efficiency and Productivity (10.9\%).} For example ``speed up the completion of tasks'' (r8), are cited as reasons for student adoption.
    \item \textbf{T6. Risks and Governance (9.1\%).} A portion of the respondents emphasize that student use must be governed by ethics and supervision to prevent plagiarism, dependence, ``because they generally or mostly use it to cheat'' (r2).
\end{itemize}
\end{comment}

Perceived benefits of students' AI use similarly focus on efficiency, where AI helps students complete work faster through improved access to information and resources. Additionally, personalization emerges as a significant theme, ranging from enabling interactive learning to assisting brainstorming.

Among the minority of respondents who were undecided or did not support AI use, we asked follow-up questions: \textbf{Q10b: ``I will support the use of AI by teachers if..."} and \textbf{Q9b: ``I will support the use of AI by students if..."}. Across both questions, responses consistently emphasized the need for monitoring, governance, and clear guardrails to prevent misuse (e.g., ``in the absence of adequate supervision, there is a risk of misuse" [r39]).

\subsubsection{Opposition and Perceived Risks}

This section explores the factors driving opposition to AI integration in education and the perceived risks respondents associated with its use. While support for AI was largely motivated by efficiency, opposition was most commonly framed around the detrimental effects of over-reliance. This over-dependency is perceived to foster complacency and skill degradation, negative outcomes that are intrinsically linked to the efficiency gains AI provides. In addition, while teachers value AI assistance in their daily activities, they expressed caution about the risk of inaccurate outputs.

To understand the specific grounds for resistance, we asked \textbf{Q11: ``I will not support AI use by teachers if...''}. Responses clustered into six themes. \textbf{(T1) Quality Control and Pedagogical Fit} (26.4\%) expressed concerns when AI is used without verification or adaptation to the classroom context (e.g., ``if used without filtering the information presented by AI'' [r28]), which may result in misleading content or misalignment with the curriculum. \textbf{(T2) Dependency and Shortcut Mentality} (24\%) reflected concerns that AI would encourage over-reliance or a ``shortcut'' approach to teaching (e.g., when AI is ``used as the primary source in learning without further developing it.'' [r36]). \textbf{(T3) Replacement of the Teacher’s Role} (11.2\%) is concerned with the risk of AI replacing the teacher’s role or reduce essential human interaction (e.g., ``it may reduce important aspects of education such as the development of social and emotional skills.'' [r19]). \textbf{(T4) Academic Integrity and Misuse} (6.4\%) focused on unethical use of AI, including for cheating or when it results in a lack of originality in teacher-produced work (e.g., ``simply copy-pasting without reading it first'' [r33]). \textbf{(T5) Harmful Content and Social Harm} (19.2\%) reflected concerns about hoaxes, hate speech, or content that violates social or religious norms (e.g., ``ssed for unethical purposes'' [r121]. Finally, \textbf{(T6) Governance and Implementation} (12.8\%) included worries about data privacy (e.g., ``AI collects private data when it is used'' [r101]), insufficient competence to use AI, and practical barriers such as costs, as reasons to not support AI use.

We observed a parallel pattern in teachers' opposition to students' use of AI. When asked \textbf{Q12: ``I will not support the use of AI by students if...''}, respondents expressed significant apprehension about skill erosion due to dependency, alongside heightened concerns about academic dishonesty. Responses clustered into four themes. \textbf{(T1) Academic Cheating and Plagiarism} (23.8\%) reflected opposition when AI is used for dishonest purposes in assignments or exams, such as ``creating research reports by copying 100\% from AI'' [r99]. \textbf{(T2) Dependency and Learning Erosion} (34.6\%) captured concerns when AI replaces the actual learning process, resulting in shallow understanding, or students becoming ``unable to think critically based on their own reasoning.'' [r24]. \textbf{(T3) Non-learning Misuse} (30.8\%) described opposition when AI is used for off-topic, negative, or unsafe purposes, such as ``seeking entertainment without age limits'' [r94]. Finally, \textbf{(T4) Conditional Support and Guardrails} (10.8\%) expressed that support depends on clear ethical guidelines, boundaries and supervision, and is withdrawn when ``its use has no clear limits.'' [r63] %or when ethical guidelines are not followed. 

\begin{comment}
    \begin{itemize}
    \item \textbf{T1. Academic Cheating and Plagiarism (23.8\%).} Opposition is noted when AI is used for dishonest purposes in assignments or exams, such as ``creating research reports by copying 100\% from AI'' (r99) .
    \item \textbf{T2. Dependency and Learning Erosion (34.6\%).} Teachers express concern when AI replaces the actual learning process, shallow understanding, or ``unable to think critically based on their own reasoning.'' (r24).
    \item \textbf{T3. Non-learning Misuse (30.8\%).} The use of AI for off-topic, negative, or unsafe purposes, such as ``seeking entertainment without age limits'' (r94), is a reason for withholding support.
    \item \textbf{T4. Conditional Support and Guardrails (10.8\%).} Support is conditional, meaning it is withdrawn if there is a lack of teacher supervision, ``Its use has no clear limits.'' (r63) or a failure to follow ethical guidelines.
\end{itemize}
\end{comment}

Overall, these findings point to a key tension: although support for AI is broad, it is tempered by substantial concerns about cognitive-skill erosion and academic integrity. Among the minority of teachers who fundamentally oppose AI use, their reasoning aligns closely with these risks. For example, one respondent noted, ``Because this can make students accustomed to artificial intelligence, making students reluctant to think critically to solve the problems given'' [r35].

\subsection{RQ.4: Pedagogical Stances of Student AI Use}

\begin{figure}[ht!]
     \centering
     \begin{subfigure}[b]{0.99\textwidth}
         \centering
         \includegraphics[width=\linewidth]{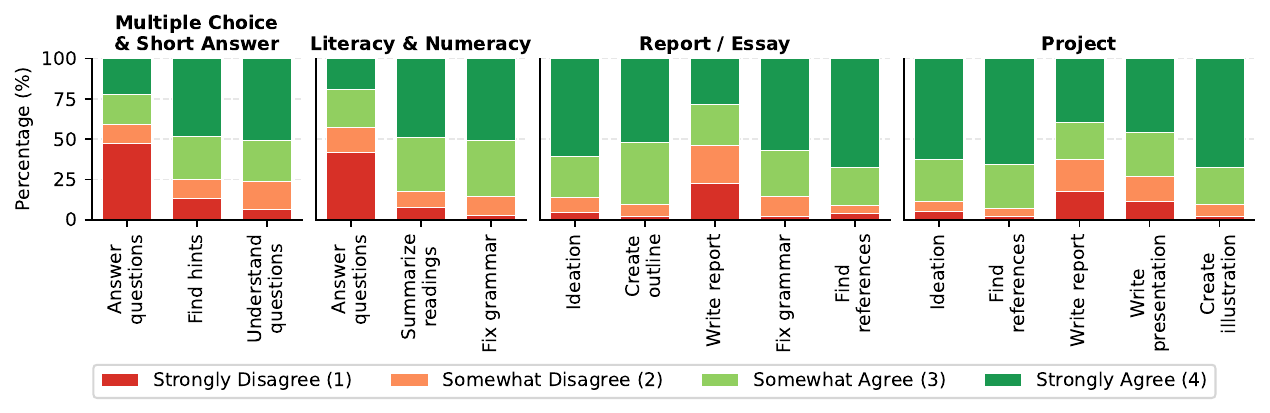}
     \end{subfigure}
     \caption{Support of AI Use for Teachers and Students.}
     \label{fig:ai_use}
     
\end{figure}

This section discusses teacher agreement regarding student AI use across various learning task, specifically: short-answer tasks, literacy/numeracy assignments, essays, and projects. We presented respondents with a multiple-response question, \textbf{Q13: ``To your knowledge, what do students use AI for in their learning activities?''}, which encompasses the four aforementioned options. For each task selected, we then employed a four-point Likert scale to assess the level of support for AI integration of activities correspond to each task in Figure~\ref{fig:ai_use}.

As discussed, teachers generally support students' use of AI. Figure \ref{fig:ai_use} further reflects this supportive stance, with a few notable exceptions: the most contested uses are using AI for answer seeking in performance-based tasks (e.g., MCQs and literacy/numeracy tasks) and AI-written reports for report-based assignments. Even so, some teachers remain strongly supportive and opposition, especially to report writing, remains a minority view. Given growing concerns about AI's educational effects, pedagogical integrity should take priority over mere efficiency: using AI to produce direct answers can undermine learning, and even ideation requires careful use to avoid harms teachers identified.

\section{Discussion}

Our findings reveal a disconnect between AI adoption and AI literacy among Indonesian educators. While AI has rapidly penetrated the teaching profession, underlying conceptual models remain superficial or misinformed. This gap creates a critical tension because teachers are increasingly reliant on a ``black box'' they lack the literacy to evaluate. This literacy lag is particularly evident in the prevalence of high level or vague descriptions of AI mechanics, which suggests that while the tools are being utilized, their internal logic remains opaque.

Paradoxically, while educators are acutely aware of systemic risks such as academic dishonesty, cognitive over-reliance, and student laziness, their support for AI integration remains remarkably high. This approval persists even when teachers perceive student usage as primarily a shortcut for direct answer seeking. Such a stance suggests that the perceived utility of AI in managing workloads currently outweighs traditional pedagogical caution. This results in a permissive environment where AI is supported across most use cases, despite a clear recognition of the potential long term harms.

\vspace{2mm}

\noindent\textbf{Interpreting AI Literacy in Context} These findings do not suggest teachers need specialist AI expertise; rather, they require the functional AI literacy to judge when AI supports learning versus when it requires verification or risks undermining the learning process. When teachers have inaccurate mental models, such as treating ChatGPT as a search engine, it negatively shapes classroom policies, monitoring practices, and the normalization of AI-supported activities.

% These results should not be read as requiring teachers to possess specialist AI expertise. In the context of this study, AI literacy refers to the functional conceptual understanding and pedagogical judgment needed to decide when AI can support learning, when its outputs require verification, and when its use risks replacing the learning process itself. For example, when teachers interpret ChatGPT as a search engine or database retrieval system, they may overestimate the reliability and completeness of its outputs and underemphasize the need for verification before using them in teaching materials, assessment support, or student guidance. The misconceptions we observed matter not because teachers lack advanced technical knowledge, but because inaccurate mental models can shape classroom rules, monitoring practices, and the kinds of AI-supported activity teachers normalize.
\vspace{2mm}

\noindent\textbf{AI-Use in Surveys} We found several respondents who used AI in the questions. In one of the questions, their answer included the phrase ``As an AI language model developed by Google,'' which shows that they did not double-check the response and simply copied and pasted it. Another respondent had most of their responses flagged as suspicious, with the remaining two questions confirmed as AI-generated. One of these included markdown formatting which serves no purpose on our survey platform. We find this result interesting and ironic, as it anecdotally illustrates the negative side effects of AI that the respondents themselves confirmed, such as over-reliance, laziness, and the need for instantaneous results, including a failure to check for correctness for the sake of work efficiency.

\vspace{2mm}

\noindent\textbf{Recommendation} Based on the misconceptions identified, we recommend a more practice-oriented approach to AI literacy and governance in Indonesian K–12 schools. Although the Ministry of Primary and Secondary Education provides initial training, many teachers still hold inaccurate mental models (e.g., treating ChatGPT as a search engine). At the same time, teachers are not unaware of risks: many explicitly mentioned harms such as dependency and dishonesty, yet some still support AI for direct answer seeking and many continue to support AI for report writing, reflecting how strongly perceived efficiency and convenience shape adoption decisions.

To close this gap, AI literacy programs should be longer and directly address misconceptions about accuracy, hallucinations, bias, and privacy~\cite{unesco_genai_guidance_2023}. Training should also require \textit{demonstrated} responsible use through authentic tasks (e.g., producing an AI-supported lesson plans, documenting verification steps, and explaining how over-reliance will be mitigated), rather than relying on self-reported readiness so that learning quality remains the priority~\cite{dfe_genai_edu,oecd_deo_2026,teachai_toolkit}. In parallel, guidance should be consolidated in a simple portal and written as short, operational documents with concrete ``do/don't" examples%with clear, concrete language. International practice shows that school-facing guidance is most effective as short, operational documents with practical examples of what to avoid
~\cite{dfe_genai_edu,nsw_genai_guidelines_2025}, supported by follow-up checks such as scenario-based assessments or micro-certifications to verify classroom-relevant competence%that measure competence through classroom practice rather than self-reported confidence
~\cite{teachai_toolkit,unesco_genai_guidance_2023}.

At the school level, clear rules are needed: teachers %Schools must also adopt clear rules. Each teacher 
should state AI policies in the syllabus %, including 
(permissible purposes, disclosure requirements, and prohibited uses~\cite{dfe_genai_edu}), and students should document their AI process % and human-in-the-loop thinking 
to protect integrity. Schools should also form AI committees to adapt national guidance to local contexts and monitor risks~\cite{teachai_toolkit}, treating AI governance as an ongoing process rather than a one-time document%. This ensures AI adoption is an ongoing governance process rather than a one-time document
~\cite{unesco_genai_guidance_2023}. Finally, given the persistent appeal of AI's convenience and the broad support for its adoption in Indonesia, even in high-risk use cases like answer seeking and report writing, teacher training in AI should %Finally, since our findings show that AI adoption is inevitable and already receives high support, the focus must 
shift from general awareness to intensive risk mitigation. %Because educators are in favor of AI use despite recognizing risks like "laziness" or dishonesty, training must provide specific technical strategies to mitigate these harms. 
Without explicit preparation for managing these trade-offs, strong support for AI may translate into classroom practices that inadvertently compromises pedagogical integrity.

%Finally, these efforts need practical support through better infrastructure and safe tools. Schools need reliable internet access so teachers can apply what they learn and use AI in real classroom conditions. They also need access to AI tools that protect student data and reduce safety risks, so responsible AI use is possible across different regions and school types.

\vspace{2mm}

\noindent\textbf{Limitations} These findings should be interpreted with several limitations. The sample is not nationally representative, with limited data from private-school teachers outside West Java. Additionally, our recruitment method may introduce selection bias toward technology-engaged educators. Furthermore, the study relies entirely on self-reported survey responses rather than observed classroom behavior. Finally, our analysis is limited to descriptive patterns and does not include inferential subgroup comparisons.

%\section{Related Work}
%- prior surveys

\section{Conclusion}

In this study, we examined the AI conception, use cases, and pedagogical stances of 132 Indonesian K–12 educators through a mixed-methods survey. We found that while adoption and support for AI are high, actual technical understanding implied from the teacher's conceptions of AI remains superficial, with widespread misconceptions regarding AI and LLM functionality. Furthermore, teachers expressed a clear awareness of the impact on student critical thinking and academic honesty, though most teacher supports AI use including by students. We suggest that future professional development and policy frameworks must move beyond general awareness toward performance-based training that directly addresses these conceptual gaps and provides concrete strategies for risk mitigation.

\bibliographystyle{splncs04}
\bibliography{biblio}

\end{document}